\documentclass[letter,superscriptaddress,twocolumn,prl,showpacs]{revtex4-1}

    \usepackage{amsmath}
    \usepackage{makeidx}
    \usepackage{amsfonts}
    \usepackage{graphicx}%
    \usepackage{color}
    \usepackage{appendix}
    \usepackage [english]{babel} 

\makeindex

\begin{document}

\title{Strongly coupled magnons and cavity microwave photons}

\author{Xufeng Zhang}
\affiliation{Department of Electrical
Engineering, Yale University, New Haven,
Connecticut 06511, USA}

\author{Chang-Ling Zou}
\affiliation{Department of Applied Physics, Yale
University, New Haven, Connecticut 06511, USA}
\affiliation{Key Lab of Quantum Information,
University of Science and Technology of China,
Hefei 230026, Anhui, People's Republic of China}

\author{Liang Jiang}
\affiliation{Department of Applied Physics, Yale
University, New Haven, Connecticut 06511, USA}

\author{Hong X. Tang}
\email{corresponding email: hong.tang@yale.edu}
\affiliation{Department of Electrical
Engineering, Yale University, New Haven,
Connecticut 06511, USA}

\date{\today}
\begin{abstract}
We realize a cavity magnon-microwave photon system in which magnetic
dipole interaction mediates strong coupling between collective motion
of large number of spins in a ferrimagnet and the microwave field
in a three-dimensional cavity. By scaling down the cavity size and
increasing number of spins, an ultrastrong coupling regime is achieved
with a cooperativity reaching 12600. Interesting dynamic features
including classical Rabi oscillation, magnetically induced transparency,
and Purcell effect are demonstrated in this highly versatile platform,
highlighting its great potential for coherent information processing.
\end{abstract}



\maketitle \emph{Introduction.---} Systems with
strong light-matter interaction have played
crucial roles in quantum
\cite{Kimble2008,Wallquist2009} and classical
information processing \cite{Spreeuw1990,
Aspelmeyer2013} as they enable coherent
information transfer between distinct physical
platforms. It is well known that systems with
large electric dipole moment can couple strongly
with the optical fields. However, the possibility
of strong light-matter interaction via magnetic
dipoles is mostly ignored. It is only recently
that Imamoglu \cite{Imamoglu2009} has pointed out
the direction to achieve strong light-matter
interaction using collective excitations of spin
ensembles, and visioned the promise of quantum
information processing in these systems. Since
then, various implementations have been proposed
and experimentally investigated. Ensembles
including ultracold atomic clouds
\cite{Verdu2009PRL_atom}, molecules
\cite{Eddins2014PRL_Molecule}, nitrogen vacancy
centers in diamond \cite{ZhuNature2011,
KuboPRL2010_CPW, AmsussPRL2011,
Kubo2011PRL_qubit, Marcos2010PRL, Ranjan2013PRL},
and ion doped crystals
\cite{SchusterPRL2010_Ruby, Probst2013PRL_RE,
Tkalcec2014Arxiv_RE} have been used to couple to
microwave resonators or even superconducting
qubits.

Magnetic materials provide a promising
alternative to achieve strong light-matter
interaction, because they have spin density many
orders of magnitude higher than dilute spin
ensembles investigated previously. For example,
Soykal \emph{et al.} \cite{Soykal2010_PRL,
Soykal2010PRB} predicted that the
nanomagnet-photon cavity can achieve strong
light-matter interaction assisted by an extremely
large number of spins in nanomagnets. In this
Letter, we realize such a hybrid system which
consists of a sphere of yttrium iron garnet (YIG,
Y$_{3}$Fe$_{5}$O$_{12}$) and a three-dimensional
(3D) microwave cavity. This new system possesses
several distinguishing advantages. Firstly, YIG
has a high spin density
($\rho_{s}=4.22\times10^{27}\mathrm{m}^{-3}$)
exceeding previous spin ensembles by orders of
magnitudes. Secondly, spin excitations in single
crystal and highly purified YIG possess very low
damping rate. Thirdly, the spin-spin interactions
through either exchange or dipolar interaction
give rise to dispersions of spin excitations
(defined modes) in YIG, which can be used for
spatial multiplexing. It is also intriguing that
there is nonlinear interaction between
excitations in the YIG, which enable nonlinear
amplification and control of magnons. For
instance, Bose-Einstein condensates of
quasi-equilibrium magnons have been realized at
room temperature \cite{Demokritov2006_BEC}.

With the proposed hybrid system, we
experimentally demonstrate the coherent coupling
between magnons (the collective spin excitation
in YIG) and microwave photons. Because of the
large spin number in YIG, strong coupling can be
achieved. Experimental demonstration has been
previously reported using a YIG thin film on a
planar superconducting microwave cavity, and a
high cooperativity of 1350 has been achieved
\cite{Huebl2013PRL_YIG}. Here, we show that by
utilizing a spherical YIG geometry and 3D
microwave cavity, our system obtains additional
advantages such as higher quality ($Q$) factors
and more uniform coupling
\cite{Paik2011PRL_3Dcavity,
Riste2013Nature_3Dcavity,
Kirchmair2013Nature_3Dcavity}. Furthermore, our
3D system is highly tunable in various
parameters, which allows us to observe
characteristic phenomena associated with distinct
parameter regimes, including the magnetically
induced transparency (MIT, the magnetic analog of
EIT, the electromagnetically induced
transparency) and the Purcell effect. Moreover,
by scaling the device dimensions, our 3D system
can enter the so-called ``ultra-strong coupling''
(USC) regime, where the coupling rate reaches a
large fraction of the oscillation frequency,
sufficient to violate the rotating-wave approximation (RWA)
\cite{Niemczyk2010, Fuchs09, Anappara2009_USC}.
Although these important features are measured in
the classical regime, our results suggest
important prospects of operating the coupled
system in quantum regime at millikelvins where
the ferromagnetic resonance (FMR) linewidth of
YIG can go down to $1.5\,\mu$T
\cite{Spencer1961PR_YIGLoss} with magnon lifetime
extended to as long as about 4 microseconds.

\begin{figure}[tp]
\begin{centering}
\includegraphics[width=1\linewidth]{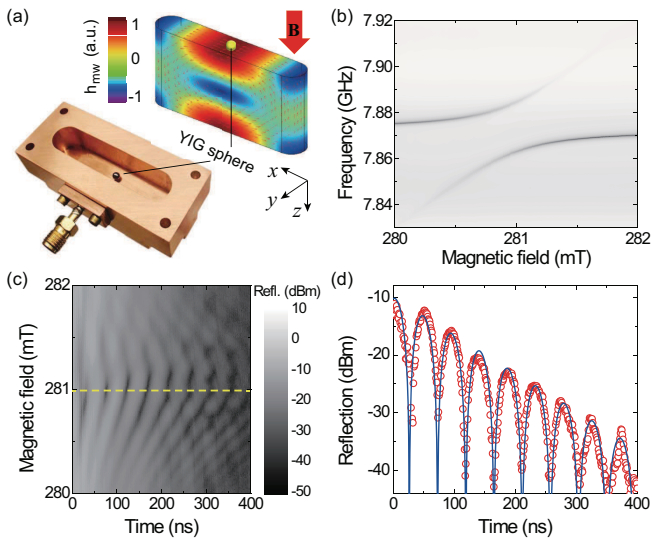}
\caption{(Color online). (a) Top: simulated
microwave cavity resonance TE101 mode
distribution. The red arrows and colors indicate
the magnetic field directions and amplitudes,
respectively. Bottom: device image showing half
of the microwave cavity with a YIG sphere inside.
(b) Measured normal mode splitting spectrum as a
function of bias magnetic field. (c) The
evolution of the cavity energy after a pulse
excitation at varying bias magnetic field. (d)
The measured Rabi oscillation signal at zero
detuned bias magnetic field. Red circles:
measurement results; Solid blue line: theoretical
calculation using parameters obtained from the
normal mode splitting spectrum.}\label{fig:sc}
\end{centering}
\end{figure}

\emph{Experimental Setup.---} The image of our
device is shown in the bottom of
Fig.\,\ref{fig:sc}(a), consisting of a 3D
microwave cavity (only the bottom half is shown)
and a highly polished YIG sphere which serves as
the magnon cavity. The microwave cavity is a box
machined from high conductivity copper to obtain
high $Q$ factor at room temperature. For example,
the box with inner dimension of
$43\times21\times9$ mm$^{3}$ gives a TE101 mode
at $\omega/2\pi=7.875$ GHz with a linewidth of a
few MHz. The simulated mode distribution (using
COMSOL 3.5) of the cavity mode is given in the
top of Fig.\,\ref{fig:sc}(a), where the arrows
and colors indicate the magnetic field directions
and their amplitudes. Spectroscopic measurement
is carried out with a vector network analyzer
(VNA) by probing the reflection of the microwave
cavity through a coaxial cable antenna.

The YIG sphere is placed inside the microwave
cavity and biased with a static magnetic field
$\overrightarrow{B_{0}}$. The magnetic components
of the microwave field perpendicular to the bias
field induce the spin flip, and thus excite the
magnon mode in YIG. Here, we are only interested
in the lowest order FMR mode, which is a uniform
collective mode that all the spins precess in
phase. This mode has the highest coupling
strength given that the microwave magnetic field
around the YIG sphere is approximately uniform
{[}Fig. 1(a){]} as the wavelength
$\lambda_{mw}\gg R$ with $R$ is the radius of the
YIG sphere. The frequency of the uniform magnon
mode linearly depends on the bias field where
$\omega_{m}=\gamma|\overrightarrow{B_{0}}|+\omega_{m,0}$,
with $\gamma=28$ GHz/T is the gyromagnetic ratio
and $\omega_{m,0}$ is determined by the
anisotropy field. The bias magnetic field is
tunable in the range of $0-2$ T, corresponding to
a magnon frequency from few hundreds of MHz to
about $50$ GHz.

\emph{Strong Coupling.---} To study the microwave
photon-magnon interaction, we adjust the bias
field so that magnon is near-resonance with the
cavity's TE101 mode. The strongest coupling
strength is obtained by placing the YIG sphere
(0.36 mm in diameter) at the position with the
maximum microwave magnetic field. The measured
microwave reflection spectra with respect to the
bias magnetic field $B_{0}$ is plotted in
Fig.\,\ref{fig:sc}(b), which exhibits an avoided
crossing at $B_{0}=281$ mT. The interaction
between microwave photon and magnon can be
described by the Hamiltonian with RWA:
\begin{equation}
\mathcal{H}/\hbar=\omega_{a}a^{\dag}a+\omega_{m}m^{\dag}m+g(a^{\dag}m+am^{\dag}),\label{eq:Hamiltonian}
\end{equation}

\noindent where $a^{\dag}$ ($a$) is the creation
(annihilation) operator for the microwave photon
at frequency $\omega_{a}$. For the magnon, the
collective spin excitations are approximately
represented by the Boson operator $m^{\dag}(m)$
with Holstein--Primakoff approximation
\cite{Holstein1940}. The coupling strength $g$
between the two systems is:
\begin{equation}
g=\frac{\eta}{2}\gamma\sqrt{\frac{\hbar\omega\mu_0}{V_{a}}}\sqrt{2Ns},\label{eq:g}
\end{equation}

\noindent where $\omega$ is the resonance
frequency and $V_{a}$ is the modal volume of the
microwave cavity resonance, $\mu_0$ is the vacuum
permeability, $N$ is the total number of spins,
and $s=\frac{5}{2}$ is the spin number of the
ground state Fe$^{3+}$ ion in YIG. The
coefficient $\eta\leq1$ describes the spatial
overlap and polarization matching conditions
between the microwave field and the magnon mode
\cite{SM}.

As shown in Fig.\,\ref{fig:sc}(b), the avoided
crossing indicates the strong coupling between
the microwave photon and the magnon, with the
coupling strength $g/2\pi=10.8\ \mathrm{MHz}$. We
can also extract the dissipation rates of both
the microwave photon ($\kappa_{a}/2\pi=2.67$ MHz)
and the magnon ($\kappa_{m}/2\pi=2.13$ MHz). The
measured spectrum agrees well with the
theoretical prediction of the reflection from the
microwave cavity:
\begin{equation}
r(\omega)=-1+\frac{2\kappa_{a,1}}{i(\omega_{a}-\omega)+\kappa_{a}+\frac{g^{2}}{i(\omega_{m}-\omega)+\kappa_{m}}},\label{eq:reflection}
\end{equation}
where $\kappa_{a,1}$ is the external coupling to
the cavity. For the coupled oscillator model
described by the Hamiltonian in
Eq.\,(\ref{eq:Hamiltonian}), hybridized
photon-magnon quasi-particles
$A_{\pm}=\sqrt{1/2}(a\pm m)$ appear for
$\omega_{a}=\omega_{m}$, with energies being
$\omega_{m}\pm g$. When the coupling strength
exceeds the dissipation rates ($g>\kappa_{a,m}$),
the system reaches the classical strong coupling
regime. With the experiment parameters, we obtain
a cooperativity of
$C=g^{2}/\kappa_{a}\kappa_{m}=21$.

The strong coupling implies coherent dynamics
between the photon and the magnon, such as Rabi
oscillations. Hence, we investigated the
temporary dynamics of photons in the strongly
coupled system. Experimentally, by monitoring the
time evolution of the cavity output after a short
pulse excitation, we obtain the time traces that
agree well with the theoretical prediction of
Rabi oscillations {[}Fig.\,\ref{fig:sc}(c){]}.
The slight asymmetry about the bias magnetic
field is due to the nonzero duration of the
excitation pulse. Clearly, the cavity energy
experiences periodic oscillation aside from the
exponential decay, demonstrating the coherent
energy exchange between photon and magnon. At
$B_0=281$ mT where the magnon is on resonance
with the microwave photon, we have the highest
signal extinction, indicating complete energy
exchange between the two systems. Also at this
bias magnetic field, the oscillation period is
the longest, corresponding to the narrowest gap
($g/\pi$) in the avoided crossing regime of the
reflection spectrum. The time trace for $B_0=281$
mT is plotted in Fig.\,\ref{fig:sc}(d), showing a
extinction ratio of more than 20 dB, and a period
of $46$ ns which agrees well with the coupling
strength $\pi/g=46.3$ ns. The calculated
oscillation signal (solid line) using the
coupling strength and the decay rate obtained
from the frequency spectrum shows excellent
agreement with the measured time trace (circles).
\begin{figure}[tp]
\begin{centering}
\includegraphics[width=1\linewidth]{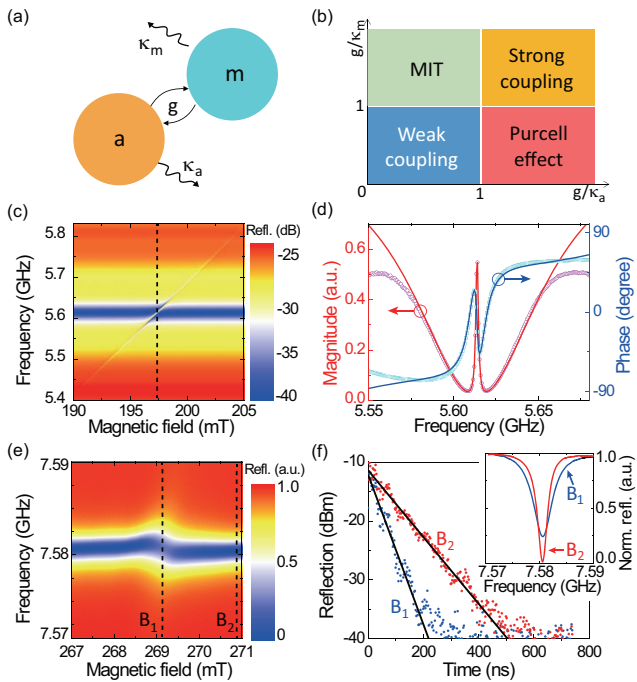}
\caption{(Color online). (a) Schematic of the
linearly coupled magnon ($m$) and photon ($a$)
system. $g,\kappa_{m},\kappa_{a}$ are the
coupling strength and dissipation rates of magnon
and microwave cavity modes, respectively. (b)
Different coupling regimes separated by the
relative strength between the coupling rate and
dissipation rates of photon and magnon subsystem.
(c) Magnetically induced transparency (MIT)
spectrum at various bias magnetic fields. (d)
Measured resonance spectrum (symbols) showing MIT
at zero detuning magnetic field. The solid lines
are the theoretical fitting using
Eq.\,(\ref{eq:reflection}). (e) Reflection
spectrum showing Purcell effect. (f) Cavity
energy ringdown curves at two different bias
magnetic fields $B_{1}$ (blue dots) and $B_{2}$
(red dots). Solid lines are the theoretical
fitting. Inset: resonance reflection spectra at
$B_{1}$ (blue line) and $B_{2}$ (red
line).}\label{fig:4quad}
\end{centering}
\end{figure}

\emph{MIT and Purcell Effects.---} Besides strong
coupling, the tunability of our proposed system
enables us to access other characteristic effects
of the coherent photon-magnon interaction.
Depending on the relative value of the coupling
strength and the dissipation rates, there are
different coupling regimes
{[}Fig.\,\ref{fig:4quad}(b){]}. We will focus on
coherent interactions with
$C=\frac{g}{\kappa_a}\cdot\frac{g}{\kappa_m}\gtrsim1$.

When the dissipation of the microwave cavity
becomes dominant in the coupled system {[}Fig.
\ref{fig:4quad}(a), $\kappa_{m}<g<\kappa_{a}${]},
the avoided crossing feature in the measured
spectrum will disappear. Indeed, as we tune the
bias magnetic field, an MIT window passes through
the broad microwave cavity resonance
{[}Fig.\,\ref{fig:4quad}(c){]}. Depending on the
detuning of the magnon frequency, the
transparency window shows up as an asymmetric
Fano-shape or symmetric peak. When the impedance
matching {[}$\kappa_{a,1}=\kappa_{a}/2$ in
Eq.\,(\ref{eq:reflection}){]} and on-resonance
conditions are satisfied, the MIT window height
is
$|r(\omega_{a})|^{2}=\left(\frac{C}{1+C}\right)^{2}$
and linewidth is $\Delta=2(1+C)\kappa_{m}$. When
the magnon is tuned on resonance with the
microwave photon [at $B_0=197.4$ mT, indicated by
the dashed line in Fig.\,\ref{fig:4quad}(c)], we
have maximum extinction with a Lorentzian-shaped
transparency window that replicates the magnon
resonance [Fig.\,\ref{fig:4quad}(d)]. The
measured transparency window has a peak height of
half unity with a maximum group delay of 110 ns.
From the measured data, the corresponding
dissipation rates and the coupling strength are
fitted using Eq.\,(\ref{eq:reflection}) as
$\kappa_{a}/2\pi=34.9$ MHz,
$\kappa_{m}/2\pi=0.24$ MHz and $g/2\pi=5.4$ MHz,
corresponding to a cooperativity value of
$C=3.76$ for this specific device configuration.

On the other hand, when the magnon decay
dominates {[}Fig.\,\ref{fig:4quad}(a),
$\kappa_{a}<g<\kappa_{m}${]}, we enter the
Purcell regime, with an enhanced decay of the
microwave cavity photon due to its coupling to
the lossy magnon. According to
Eq.\,(\ref{eq:reflection}), the effective
dissipation rate of the cavity is
$\kappa_{a}(1+C)$, enhanced by a Purcell factor
($F_{\mathrm{P}}=C+1$) as a result of the
photon-magnon interaction. Such linewidth
broadening is confirmed by the measured
reflection spectra at various bias magnetic
fields {[}Fig.\,\ref{fig:4quad}(e){]}. Although
the magnon resonance cannot be resolved due to
its large linewidth, its magnetic dependence is
inherited by the coupled mode and shows up as a
small bend. For a clear comparison, the resonance
spectra of the microwave cavity with (at
$B_{1}=269.2$ mT) and without (at $B_{2}=270.8$
mT) the coupling to the magnon are plotted in the
inset of Fig.\,\ref{fig:4quad}(f). Due to the
Purcell effect, the dissipation rate of the
microwave resonance ($\kappa_{a}/2\pi$) increases
from $1.07$ MHz to $2.09$ MHz, and the extinction
ratio is reduced by 25\%. Thus, we have
$F_{\mathrm{P}}=1.95$ and $C=0.95$. From the
experiment results, we can extract the coupling
strength $g/2\pi=3.1$ MHz and the magnon decay
rate $\kappa_{m}/2\pi=19$ MHz that indeed fall
inside the Purcell regime.

A more direct characterization of such Purcell effect is obtained
by measuring the cavity photon lifetime. Since now the magnon dissipates
very quickly, an accelerated exponential decay instead of a Rabi oscillation
is expected after a pulsed excitation. The decay curves at bias magnetic
fields $B_{1}$ and $B_{2}$ are plotted in Fig.\,\ref{fig:4quad}(f),
which gives a lifetime of $\tau_{1}=33.4\pm5$ ns and $\tau_{2}=69.8\pm7$
ns, respectively. These time domain measurements perfectly match the
dissipation rates measured above in the frequency domain. Both the
time and frequency domain measurements give a Purcell factor of about
$2$.

\begin{figure}[tp]
\begin{centering}
\includegraphics[width=1\linewidth]{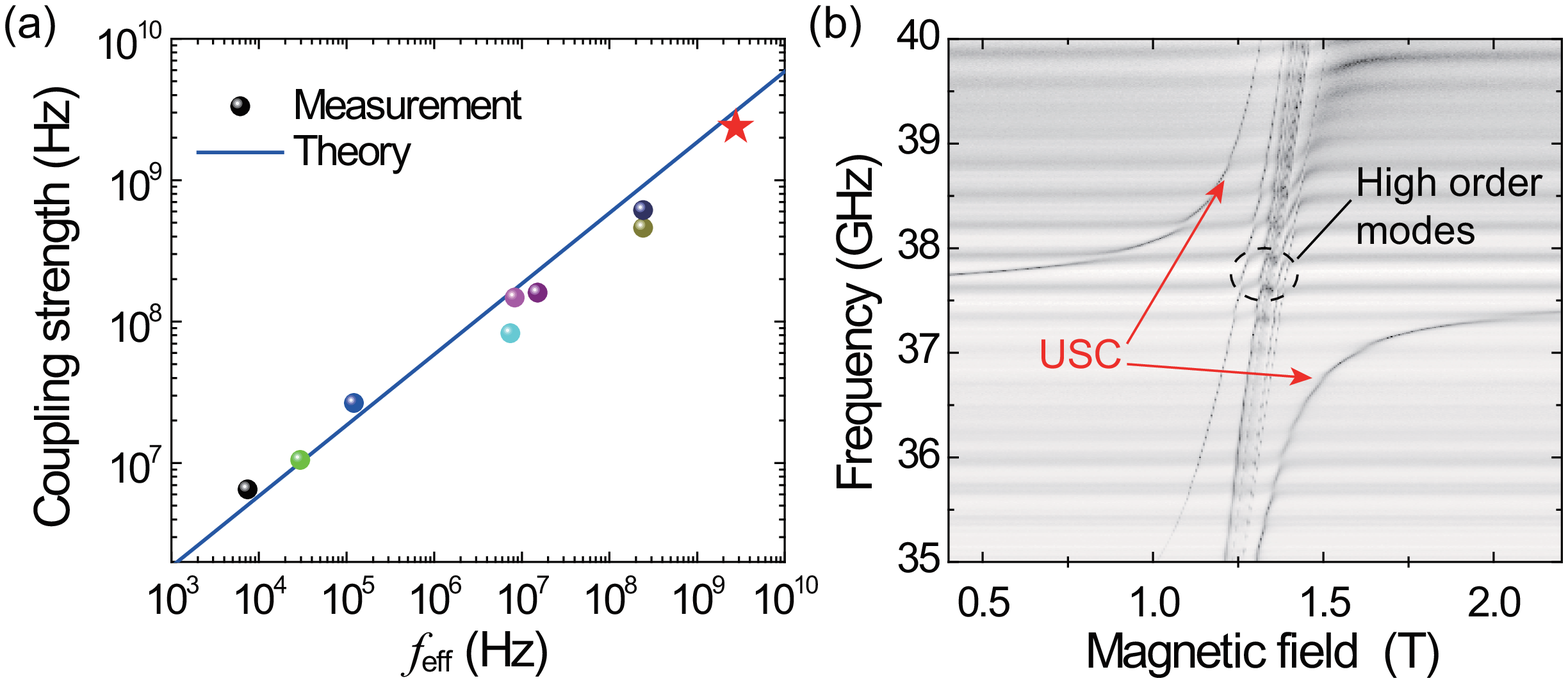}
\caption{(Color online). (a) Coupling strength as
a function of modal frequency $f_{\mathrm{eff}}$.
Solid line is the theoretical prediction. Red
star indicates a device reaching the ultrastrong
coupling. (b) Ultrastrong coupling spectrum in
the $\mathrm{K}_{\mathrm{a}}$ band. Red arrows
show the ultrastrongly coupled magnon-microwave
photon mode.}\label{fig:ultra}
\end{centering}
\end{figure}

\emph{Ultrastrong Coupling.---} Beyond the four
coupling regimes discussed above determined by
the ratio of coupling strength and dissipation
rates, there exists a USC regime where the
coupling strength becomes considerably comparable
with the magnon frequency. In the USC regime, the
RWA used in the Hamiltonian given by Eq.
(\ref{eq:Hamiltonian}) is no longer valid. The
USC has attracted intensive interests being a
potential playground for ultra-fast coherent
controlling and exploring new physics beyond RWA.
It is notable that in our experiments by
engineering the microwave cavity and the YIG
sphere, we can \emph{tune} the coupling strength
and eventually extend our coupled system to the
USC regime. From Eq.\,(\ref{eq:g}) we can see
that the coupling strength $g\propto
f_{\mathrm{eff}}=f\frac{V_{m}}{V_{a}}$, where
$V_{m}$ is the volume of the YIG sphere which
determines the spin number ($N\propto V_{m}$). By
increasing the resonance frequency and the YIG
sphere size while reducing the microwave cavity
size we can increase $f_{\mathrm{eff}}$ and
consequently the coupling strength $g$. The
coupling strengths measured in nine devices of
varying cavity and sphere dimensions are
displayed in Fig.\,\ref{fig:ultra}(a) as a
function of $f_{\mathrm{eff}}$. Good agreement is
obtained in comparison with the theoretical
prediction (solid line). During the experiments,
different resonance frequencies ranging from the
X band to the K$_{\mathrm{a}}$ band ($7$ GHz to
$40$ GHz) are tested, showing the great
tunability of the magnon.

An ultra-strong coupling strength of $g/2\pi=2.5$
GHz (represented by the star) has been achieved
at a resonance frequency of
$\omega_{a}/2\pi=37.5$ GHz, where the microwave
cavity size is dramatically reduced to
$7.0\times5.0\times3.2$ mm$^{3}$, and the YIG
sphere diameter is increased to $2.5$ mm
(corresponding to 3.5$\times 10^{19}$ spins).
This coupled system yields a ratio of
$g/\omega_{a}=6.7\%$ and reaches the USC regime.
Figure\,\ref{fig:ultra}(b) plots the reflection
spectrum of the USC. Due to the large YIG sphere
size which is now comparable with the microwave
cavity, the microwave fields penetrating the YIG
sphere is no longer as uniform and therefore
excites non-uniform magnon modes, as labeled in
Fig.\,\ref{fig:ultra}(b). These non-uniform modes
have higher frequency, and mostly couple weakly
with the microwave cavity. Also at such high
frequencies, the cavity resonance experiences
higher losses. Nevertheless, due to the ultrahigh
coupling rate, an ultra high cooperativity of
$C\simeq12600$ is realized with the extracted
dissipation rates of the microwave photon and the
magnon resonance are $\kappa_{a}/2\pi=33$ MHz and
$\kappa_{m}/2\pi=15$ MHz, respectively.

\emph{Conclusion.---} We have experimentally
realized coherent coupling between microwave
photon and magnon at room temperature, and
demonstrated the great potential of magnon as an
information carrier. Strong coupling with high
cooperativity has been achieved using
spectroscopic measurement, and the coherent
energy exchange has been illustrated with the
Rabi oscillation measurement in the time domain.
Both the MIT and Purcell effects have been
observed, providing various possible applications
for our proposed system. The coherent coupling
can be further extended into the USC regime with
a coupling strength of 2.5 GHz measured at 37.5
GHz carrier frequency, which is a new regime
where the RWA approximation may be invalidated.
Compared with previous systems, our system
possesses a range of advantages thanks to the
collective motion of large number of spins and
uniform magnetic coupling. The excellent
tunability properties of magnon system, together
with its extended lifetime, reduced thermal
excitation and the ability of coupling to
microwave qubits, makes it a very promising
candidate as a transducer that can interconnect
different systems such as photonics, mechanics
and microwave circuits.

\emph{Note added.---}While we are preparing the
manuscript, another interesting work by Tabuchi
\emph{et al}. on strongly coupled YIG-microwave
cavity has appeared on arxiv \cite{Tabuchi2014}.

\begin{acknowledgements} This work is supported by DARPA/MTO MESO
program. H. X. T. acknowledges support from a Packard Fellowship in
Science and Engineering. LJ acknowledges
support from the Alfred P Sloan Foundation, the Packard Foundation, and the DARPA
Quiness program. The authors thank Dr. Michel H. Devoret for
providing a prototype 3D microwave cavity. \end{acknowledgements}

\appendix
\section{Appendix A: Coupling strength and its dependence on the sphere location}

\begin{figure}[htpb]
\begin{centering}
\includegraphics[width=1\linewidth]{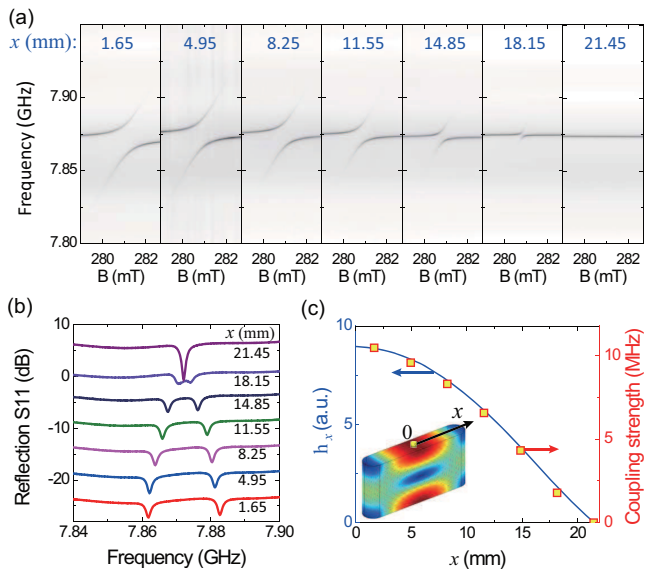} \caption{(Color online). (a) Reflection spectra at various positions as a
function of frequency and bias magnetic field.
(b) Reflection spectra at zero detuning magnetic
as the sphere is moved away from the center. (c)
Coupling strength (measured, red squares) and
microwave field amplitude (simulated, solid blue
line) as a function of
position.}\label{fig:position}
\end{centering}
\end{figure}

The overlapping coefficient $\eta$ in Eq.\,(2) in
the main text can be explicitly written as:
\begin{equation}
\label{eq:eta}
\eta^{2}=\frac{(\overrightarrow{h}(\mathbf{r})\cdot\overrightarrow{e_{x}})^{2}+(\overrightarrow{h}(\mathbf{r})\cdot\overrightarrow{e_{y}})^{2}}{\mathrm{max}\{|\overrightarrow{h}(\mathbf{r})|^{2}\}},
\end{equation}
\noindent where
$\mathrm{max}\{|\overrightarrow{h}(\mathbf{r})|^{2}\}$
is the maximum magnetic field intensity of the
cavity mode, and $\overrightarrow{h}(\mathbf{r})$
is the magnetic field amplitude at the location
$\mathbf{r}$ of the YIG sphere,
$\overrightarrow{e_{j}}$ with $j=x,y,z$ are unit
vectors and $\overrightarrow{e_{z}}$ is along the
bias field direction. In our experiment, the YIG
sphere is attached to the cavity wall and moved
only along the $x$ direction ($x_{max}=21.5$ mm
at the cavity boundary), as shown in the inset of
Fig.\,\ref{fig:position}(c). Due to the
$x$-dependent spatial distribution of the ac
magnetic field within the microwave cavity, the
coupling strength between the magnon and the
microwave photon ($g\propto\eta$) can be adjusted
by placing the YIG sphere at different position
$x$. In addition, the fields near the wall is
polarized along $x$-direction, thus we have
\begin{equation}
\label{eq:eta_approx}
\eta\approx\frac{\overrightarrow{h}(\mathbf{r})\cdot\overrightarrow{e_{x}}}{\mathrm{max}\{\overrightarrow{h}(\mathbf{r})\cdot\overrightarrow{e_{x}}\}}.
\end{equation}

Figure\,\ref{fig:position}(a) shows the
reflection spectra at various position $x$ as a
function of frequency and the bias magnetic
field. The gap of the avoid crossing becomes
smaller as we move the YIG sphere away from the
magnetic field maximum ($x=0$). The spectra at
zero detuning magnetic field, where the magnon is
on resonance with the microwave photon, is
plotted in Fig.\,\ref{fig:position}(b). From
these spectra, the coupling strength is extracted
as the half of the frequency split.
Figure\,\ref{fig:position}(c) shows the extracted
coupling strength as a function of $x$ (red
squares), and is in good agreement to the
distribution of the microwave magnetic field
(obtained by numerical simulation using COMSOL
3.5) according to Eq.\,(\ref{eq:eta_approx})
(solid blue line).

\section{Appendix B: Linewidth engineering}

In order to reach different coupling regimes, the
linewidths of both the microwave and the magnon
cavities are adjusted. The highest quality ($Q$)
factors for both cavities are determined by the
absorption of the material, thus it is
impractical to further increase the them. To
adjust the $Q$ factor of the microwave resonance,
we use a piece of microwave absorber to increase
the absorption loss. To adjust the $Q$ factor of
the magnon, iron filings are glued on the YIG
sphere surface to increase the absorption and
scattering losses. In both cases, the $Q$ factors
can be varied from the intrinsic value (several
thousands) to below 100, allowing us to explore
different coupling regimes.

\end{document}